# The transmembrane potential across a charged nanochannel subjected to asymmetric electrolytes

Ramadan Abu-Rjal and Yoav Green*
Department of Mechanical Engineering, Ben-Gurion University of the Negev, Beer-Sheva 8410501, Israel

* Email: yoavgreen@bgu.ac.il
ORCID numbers: R.A-R.: 0000-0002-1534-9710
Y.G. : 0000-0002-0809-6575

The transmembrane voltage, $V$, which is the potential drop required to nullify the electrical current ($i$=0), is a key characteristic of water desalination and energy harvesting systems that utilize macroscopically large nanoporous membranes, as well as for physiological ion channels subjected to asymmetric salt concentrations. To date, existing analytical expressions for $V_{i=0}$ have been limited to simple scenarios and/or under simplifying assumptions. In this work, we derive two expressions for $V_{i=0}$. First, we consider the much simpler scenario of two species. Then, we can consider an electrolyte comprised of an arbitrary number of species. The difference in the models is that the latter solution utilizes an ad-hoc assumption of a linear concentration profile, while the former solution does not require such an ad-hoc assumption. We analyze both models and show how to reduce them to several known models. We also verify both models with numerical simulations of the one-dimensional Poisson–Nernst–Planck equations. We show how the interplay between diffusion coefficients and ionic valencies significantly varies the system response and why it is essential to account for all system parameters. Ultimately, this model can be used to improve experimental interpretation of ion transport measurements.

## I. INTRODUCTION

Ion transport through nanoscale charged channels is ubiquitous in both technological applications and natural systems. In technologies such as water desalination [1–4] and energy harvesting [5–9] via electrodialysis and reverse-electrodialysis processes, respectively, large-scale nanoporous membranes are employed. In biological systems, the channels are protein ion channels that regulate basic physiological phenomena [10–13], while in biomolecule sensing [14–20] and DNA sequencing [19,21–27] systems, nanopores are utilized.

Although these systems often differ in geometry (e.g., length scales), materials (e.g., surface charge distribution), and electrolyte (concentrations, diffusion, and more), they all exhibit a universal transport behavior. The universality of the response can be attributed to the fact that, in all these scenarios, regardless of their macroscopic and microscopic structure, ion transport is governed by the same underlying electrokinetic principles, allowing insights gained in one system to be applicable to others.

To elucidate this universality, it is essential to identify the features common to all such systems. Figure 1 schematically illustrates a generic configuration in which a charged channel connects two large electrolyte reservoirs with asymmetric salt concentrations, wherein the entire system is subjected to a potential drop, $V$. The channel may represent a nanochannel, a nanopore, a nanoporous membrane, or a biological ion channel; for clarity, throughout this work, we will refer to it as a “nanochannel” or, in short, “channel”.

When the salt concentrations differ at the two ends of the nanochannel, a potential difference $V$ develops across the channel. In the absence of current ($i$ = 0), this voltage ($V_{i=0}$) – known variously as the resting, open-circuit, osmotic [28,29], or reversal [30,31] voltage or transmembrane potential [32,33] – serves as a key characteristic of these systems. This voltage, for example, sets the minimum voltage required for ion transport in the electrodialysis process. In physiology, it represents the electric potential barrier that needs to be overcome for physiological events to occur. The $V_{i=0}$ has therefore long attracted significant attention, and for more than a century, numerous works have sought to develop theoretical descriptions of the $V_{i=0}$ across biological and artificial membranes [34–43]. However, existing analytical expressions for $V_{i=0}$ have been limited to simplified scenarios, such as symmetric univalent salts, specific concentration gradients or electrolytes, zero surface charge density, or an incorrect assumption of a uniform electric field.

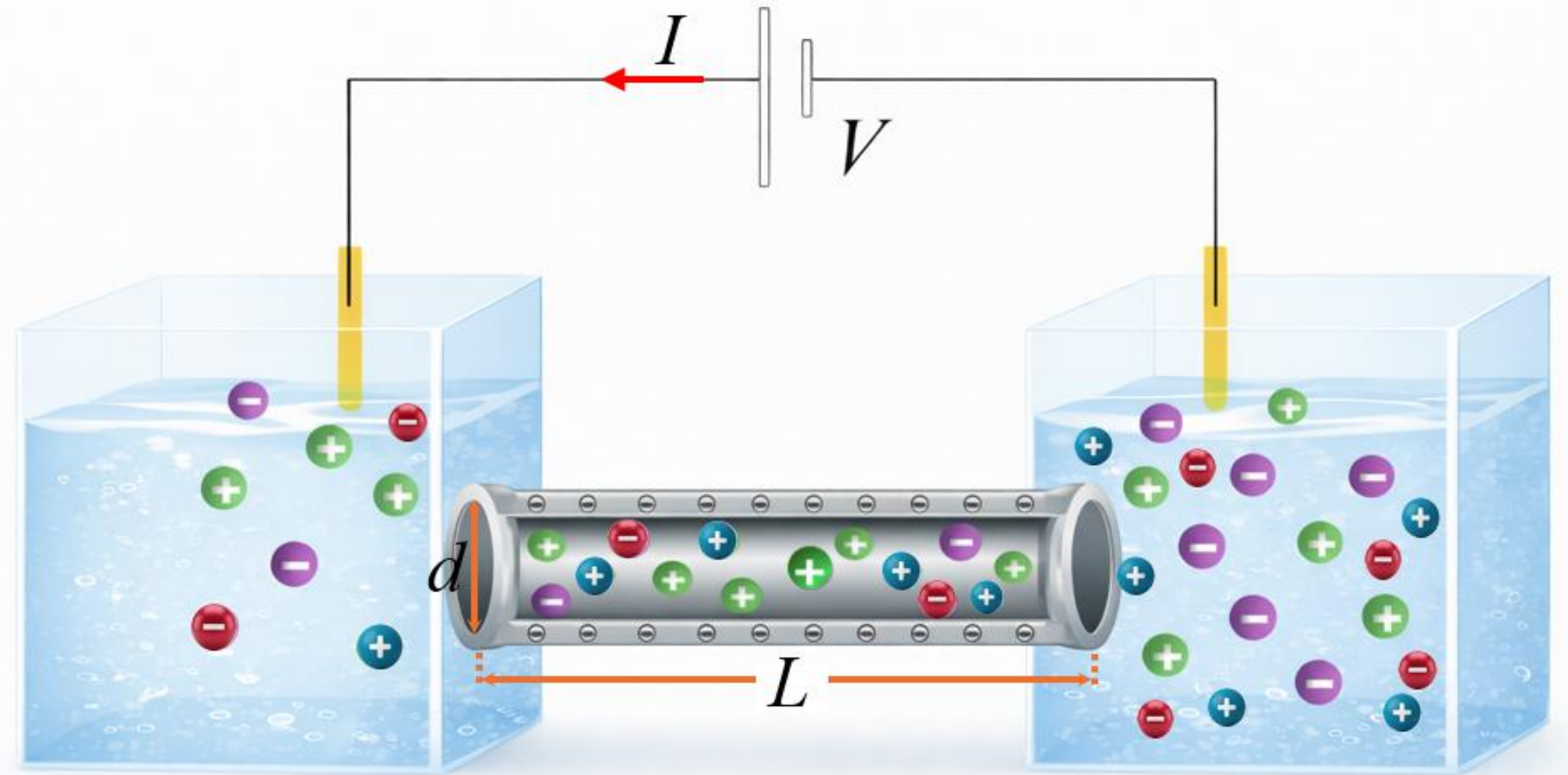

**Figure 1.** Schematic of a nanofluidic system subjected to a voltage drop $V$ (defined as positive from left to right) and asymmetric salt concentrations, $c_{\mathrm{left}}$ and $c_{\mathrm{right}}$, at the two ends of the nanochannels. The surface charge density, which is defined here to be negative, results in an excess of positive ions, represented by green and blue spheres, over the negative ions, represented by purple and red spheres.

In this work, using the classical continuum-based Poisson-Nernst-Planck equations, subject to the assumption of electroneutrality, we derive two models for the zero-current voltage, $V_{i=0}$. In Sec. II, we present the model. In particular, we discuss the geometry, governing equations, and boundary conditions for an arbitrary number of species. In Sec. III, we derive the solution for two species. Here, we introduce the method for calculating the Donnan potential drop. For two species, we show that the potential drop can be analytically evaluated for several scenarios. In Sec. IV, we expand this methodology to hold for more species, and show that one needs to numerically calculate the Donnan potentials using an analytically derived set of equations. Both in Secs. III and IV, we show how to reduce the new models to several previous models, as well as we present results for each of the scenarios. In Sec. V, we conclude with a discussion on the results of this work and how this work can be used to address several open questions.

## II. MODEL

### A. System setup and governing equations

We consider a charged nanochannel of length $L$, cross-sectional area $A$, and perimeter $P$. Figure 1 shows a cylindrical channel; however, generally, this need not be the case. The nanochannel is charged with a uniform surface charge density $\sigma_s$, and filled with an electrolyte comprising an arbitrary number, $K$, of species, some of which are positive while others (at least one) are negative. The properties of each species will be denoted with the index $k$.

The transport of the species is governed by the Poisson-Nernst-Planck (PNP) equations. In steady state, under the assumption of negligible advection, the one-dimensional (1D) equations can be written

$$-j_k = D_k \partial_x c_k + V_{\text{th}}^{-1} D_k z_k c_k \partial_x \phi \,, \tag{1}$$

$$\varepsilon_0 \varepsilon_r \partial_{xx} \phi = -\rho_e = -F \sum_{k=1}^{K} z_k c_k + F \Sigma_s \,. \tag{2}$$

Equation (1) is the Nernst-Planck equation for conservation of flux, $j_k$, of species $c_k$, where $D_k$ are the diffusion coefficients, $z_k$ are the valences , and $V_{\text{th}} = RT / F$ is the thermal voltage, which depends on the universal gas constant, $R$, the absolute temperature, $T$, and the Faraday constant, $F$. Equation (2) is the Poisson equation for the electric potential, $\phi$, which depends on the space charge density, $\rho_e$, the permittivity of free space, $\varepsilon_0$, and the relative permittivity, $\varepsilon_r$. Derivatives in the $x$ direction are denoted by $\partial_x$. The space charge density, $\rho_e$, is the valence-weighted sum of the concentrations [Eq. (2)]. The excess counterion concentration, $\Sigma_s$, due to the surface charge, is the perimeter integration of the space charge density after it has been divided by the cross-sectional area (effectively reducing the 3D problem to a 1D formulation) [42–45]

$$\Sigma_s = -\sigma_s P / FA \,. \tag{3}$$

The system is subjected to a potential drop, $V$, and asymmetric bulk concentrations

$$\phi(x=0) = V \;\;,\;\; \phi(x=L) = 0 \;\;,\;\; c_k(x=0) = c_{k,\text{left}} \;\;,\;\; c_k(x=L) = c_{k,\text{right}} \,. \tag{4}$$

### B. The total potential drop

Following previous works [39–43,45–47], the total potential drop across the system is the sum of two Donnan potential drops (the drop over the thin interfacial electrical double layers at the two interfaces at $x = 0, L$) and the potential drop across the nanochannel itself (the Ohmic drop), such that the total potential drop is then given by

$$-V = \Delta\phi_{\text{Donnan-left}} + \Delta\phi_{\text{nano}} + \Delta\phi_{\text{Donnan-right}} \tag{5}$$

In each of the sections below, we will show how to calculate the Donnan potential drops, analytically or semi-analytically. Similarly, we will also calculate the second term (the Ohmic drop) using two different approaches.

To calculate the Donnan potential, it is necessary first to define the electrochemical (EC) potential

$$\mu_k = RT \ln c_k + z_k F \phi + \mu_{\text{ref},,k} \,. \tag{6}$$

Note that the fluxes in Eq. (1) can be written in terms of the gradients of the EC potentials as

$$-j_k = (D_k / RT) c_k \partial_x \mu_k \,. \tag{7}$$

If the fluxes are continuous at the $x = 0, L$ interfaces, so are the EC potentials. As such, we shall require continuity of the EC potential at the interfaces, allowing us to ignore the reference potential, $\mu_{ref,k}$, (which will be dropped henceforth). In principle, if the EC potential is continuous at the interfaces, then one can write

$$(RT \ln c_k + z_k F \phi)_{\text{inside}} = (RT \ln c_k + z_k F \phi)_{\text{outside}} \,, \tag{8}$$

such that

$$\Delta\phi_{\text{interface}} = \phi_{\text{inside}} - \phi_{\text{outside}} = \frac{V_{\text{th}}}{z_k} \ln\left( \frac{c_{k,\text{outside}}}{c_{k,\text{inside}}} \right). \tag{9}$$

Several comments regarding Eq. (9) are warranted. First, this interfacial potential difference is related to the Donnan potential up to a plus/minus sign, which is determined by whether the inside is farther along the x-axis than the outside. Second, this equation holds for any/all the species simultaneously. Hence, whether for two species or for more, one can choose any one of the species for this calculation. Finally, in the following, we denote all the (currently unknown) internal concentrations with an overbar, while the outside concentrations are known via Eq. (4).

### C. Local electroneutrality

Since the governing equations are nonlinear and coupled, we resort to assuming that within the channel, the electrolyte is locally (and globally) electroneutral, such that $\rho_e = 0$. This assumption is allowed so long as the electrical double layers are sufficiently small relative to the system's characteristic length (i.e., the system's length) such that the Laplacian term in Eq. (2) can be taken to zero [48,49]. Then, $\rho_e = 0$ yields

$$\rho_e = F\sum_{k=1}^{K} z_k c_k - F\Sigma_s = 0 . \tag{10}$$

Note that while the local electroneutrality assumption will soon reduce the governing equations, this framework remains a formal solution of the PNP equations. It is also essential to note that in the reservoirs one needs take $\Sigma_s = 0$ while still requiring that $\rho_e = 0$. In other words, when the boundary conditions for the concentrations are prescribed in Eq. (4), one still needs to ensure that $\sum_{k=1}^{K}(z_k c_k) = 0$ .

## III. TWO SPECIES SOLUTION

In this section, we consider a binary electrolyte consisting of one positive (+) and one negative (–) species. In this case, instead of using the index *k*, the properties of each species are denoted with the respective +/– subscript. Here, the concentrations of each species can be related to the total salt concentrations in the reservoirs $c_{\text{left/right}}$, wherein the concentration in Eq. (4) can be simplified to

$$c_{\pm}(x=0) = c_{\text{left}} / |z_{\pm}| \;\; , \;\; c_{\pm}(x=L) = c_{\text{right}} / |z_{\pm}| . \tag{11}$$

### A. Derivation

#### *1. Donnan potentials*

At the $x = 0, L$ interfaces, we shall require that the EC potential be continuous inside and outside the channel. The Donnan potential drop at $x = 0$ is defined as the potential difference between the inside and the outside of the channel. Using the EC potential [Eq. (9)] for the positive species yields

$$\Delta\phi_{\text{Donnan-left}} = \frac{V_{\text{th}}}{z_+}\ln\left(\frac{c_{+,\text{left}}}{z_+\bar{c}_{+,\text{left}}}\right). \tag{12}$$

Here $\bar{c}_{+,\text{left}}$ is the still unknown concentration inside the left end of the pore. Similarly, the Donnan potential at $x = L$ is defined as the potential difference between the outside and the inside of the channel

$$\Delta\phi_{\text{Donnan-right}} = \frac{V_{\text{th}}}{z_+}\ln\left(\frac{z_+\bar{c}_{+,\text{right}}}{c_{+,\text{right}}}\right), \tag{13}$$

where $\bar{c}_{+,\text{right}}$ is the unknown concentration inside the right end of the channel.

The concentrations inside the channel can be evaluated using the EC potential, Eq. (6). We remove the EC potential's dependence on the electric potential by considering the following EC potential, which accounts for both contributions

$$z_+\mu_- - z_-\mu_+ = RT\ln(c_-^{z_+} / c_+^{z_-}). \tag{14}$$

Removing the ln dependence in Eq. (14), and requiring continuity inside and outside the channels, yields

$$(c_-^{z_+} / c_+^{z_-})_{\text{outside}} = (c_-^{z_+} / c_+^{z_-})_{\text{inside}}. \tag{15}$$

Accounting for electroneutrality [Eq. (10)], and that the outer concentrations are given [Eq. (11)], yields

$$\left(\frac{c_{\text{left/right}}}{-z_-}\right)^{z_+}\left(\frac{c_{\text{left/right}}}{z_+}\right)^{-z_-} = \bar{c}_{-,\text{left/right}}^{z_+}\left(\frac{-z_-\bar{c}_{-,\text{left/right}}+\Sigma_s}{z_+}\right)^{-z_-}. \tag{16}$$

In principle, this equation needs to be solved numerically, which can be done so rather easily with a simple Newton-Raphson solver. However, there are several simple, analytically tractable solutions. For $z_+ = -z_-$

$$\bar{c}_{-,\text{left/right}} = \frac{1}{2z_+}\left(-\Sigma_s + \sqrt{4c_{\text{left/right}}^2 + \Sigma_s^2}\right). \tag{17}$$

The expressions for $z_+/|z_-| = 1/3, 1/2, 2, 3$ are provided in the Supplemental Material (SM) [50]. Regardless, from this point, whether one has an analytical expression or a numerical value for $\bar{c}_{+,\text{left}}$ and $\bar{c}_{+,\text{right}}$, they can be considered to be known, such that the Donnan potential drop in Eqs. (12) and (13) [along with Eq. (10)] can be evaluated.

### 2. *The nanochannel potential drop,* $\Delta\phi_{\text{nano}}$

We finally return to calculating $\Delta\phi_{\text{nano}}$ in Eq. (5). We turn to calculating the electrical current

$$i = F(z_+j_+ + z_-j_-). \tag{18}$$

Inserting electroneutrality [Eq. (10)] into this equation yields [and Eq. (1)],

$$-V_{\text{th}}\frac{i}{F} = -V_{\text{th}}(D_+ - D_-)z_-\partial_x c_- + [-z_-(D_+z_+ - D_-z_-)c_- + D_+z_+\Sigma_s]\partial_x\phi. \tag{19}$$

When the electrical current is zero ($i = 0$), Eq. (19) yields

$$\frac{\partial_x\phi}{V_{\text{th}}} = \frac{z_-(D_+ - D_-)\partial_x c_-}{-z_-(D_+z_+ - D_-z_-)c_- + z_+D_+\Sigma_s}, \tag{20}$$

which can be integrated between the two inner ends of the channels, where $\bar{c}_{-,\text{left}}$ and $\bar{c}_{-,\text{right}}$ are already known. Then, we find that

$$\frac{\Delta\phi_{\text{nano}}}{V_{\text{th}}} = \frac{D_- - D_+}{z_+D_+ - z_-D_-}\ln\left[\frac{-z_-(D_+z_+ - D_-z_-)\bar{c}_{-,\text{right}} + z_+D_+\Sigma_s}{-z_-(D_+z_+ - D_-z_-)\bar{c}_{-,\text{left}} + z_+D_+\Sigma_s}\right]. \tag{21}$$

### 3. $V_{i=0}$ *for two species*

Finally, upon inserting Eqs. (12),(13) and (21) into Eq. (5), we find that the zero-current/transmembrane potential

$$V_{i=0} = \frac{V_{\text{th}}}{z_+}\ln\left(\frac{c_{\text{right}}}{c_{\text{left}}}\frac{\bar{c}_{+,\text{left}}}{\bar{c}_{+,\text{right}}}\right) + V_{\text{th}}\frac{D_+ - D_-}{D_+z_+ - D_-z_-}\ln\left[\frac{z_-(D_+z_+ - D_-z_-)\bar{c}_{-,\text{right}} - D_+z_+\Sigma_s}{z_-(D_+z_+ - D_-z_-)\bar{c}_{-,\text{left}} - D_+z_+\Sigma_s}\right]. \tag{22}$$

As we will later show, Eq. (22) is a limiting scenario for the multispecies solution that will be derived in the next section. However, in the upcoming derivation, an ad hoc assumption that is not needed here is used. Thus, to a certain extent, for two species, the derivation used here supersedes the derivation used later. Yet, we emphasize that the approach used here is limited to two species.

### B. Reduction to previous models

Equation (22) for binary salts is one of the key findings of this work. Before analyzing its behavior, it is essential to note that this equation can be reduced to several previously derived models, implying that Eq. (22) is their unifying model. The case of a $z_+ = -z_-$ salt, where the internal interfacial concentrations are given by Eq. (17), reduces $V_{i=0}$ to

$$V_{i=0} = \frac{V_{\text{th}}}{z_+} \ln\left( \frac{c_{\text{right}}}{c_{\text{left}}} \frac{\overline{c}_{+,\text{left}}}{\overline{c}_{+,\text{right}}} \right) + \frac{V_{\text{th}}}{z_+} \frac{D_+ - D_-}{D_+ + D_-} \ln\left[ \frac{z_+(D_+ + D_-)\overline{c}_{-,\text{right}} + D_+\Sigma_s}{z_+(D_+ + D_-)\overline{c}_{-,\text{left}} + D_+\Sigma_s} \right], \tag{23}$$

derived initially by Galama et al. [41], and later by Green [43,45]. Inserting $D_+ = D_-$ into Eq. (23), yields $V_{i=0}$ derived in Ref. [42] [this is merely the first term in Eq. (23)].

Equation (22) has two additional important limits: very large and very small ratios of $\Sigma_s$ to the bulk concentrations, $\Sigma_s / c_{\text{left,right}}$. For small ratios such that one can take $\Sigma_s = 0$, the channel is effectively nonselective, and Eq. (22) is drastically simplified. Here, Eq. (16) yields the trivial solution that $\overline{c}_{\pm,\text{ left}} = c_{\text{left}} / |z_\pm|$ and $\overline{c}_{\pm,\text{ right}} = c_{\text{right}} / |z_\pm|$, implying that the Donnan potential drops, given by the first term, vanish. Thus, Eq. (22) reduces to Henderson's equation [51,52]

$$V_{i=0} = V_{\text{th}} \frac{D_+ - D_-}{z_+ D_+ - z_- D_-} \ln\left( \frac{c_{\text{right}}}{c_{\text{left}}} \right). \tag{24}$$

One interesting peculiarity of Henderson's equation, which will shortly be demonstrated, is that for a given concentration ratio, the ratio of $D_+ / D_-$ determines whether $V_{i=0}$ is positive, negative, or zero. Also, note that the denominator is always positive.

For the scenario that $\Sigma_s / c_{\text{left,right}}$ is very large, i.e., a highly selective channel, the negative ions are almost excluded from the channel, such that the second term in Eq. (22) goes to zero, while the positive ion concentration is almost constant, such that the ratio $\overline{c}_{+,\text{left}} / \overline{c}_{+,\text{right}}$ goes to unity. As a result, Eq. (22) reduces to the well-known Nernst's potential [53]

$$V_{i=0} = \frac{V_{\text{th}}}{z_+} \ln\left( \frac{c_{\text{right}}}{c_{\text{left}}} \right). \tag{25}$$

Here, it is interesting to note that $V_{i=0}$ is independent of the diffusion coefficients but decreases with increasing valency of the positive ion. Note that if we had considered a channel that was positively charged (instead of our negatively charged channel) and highly selective, then Eq. (25) would be modified such that $z_+$ would be replaced by $z_-$.

### C. Results

Figure 2(a) presents the results for $z_+ = -z_-$ [Eq. (23)]. These results were discussed thoroughly in our recent works [43,45], and thus, we present them here briefly. This analysis serves as the basis for the analysis of the $z_+ \neq -z_-$ shown in Figure 2(b-c), and in the later section that analyses multiple species. Note that in all these figures, it can be seen that Eq. (22) [or Eq. (23)] shows remarkable correspondence to numerical simulations, whose details can be found in Appendix A, for all concentrations and the considered salts given in Table 2.

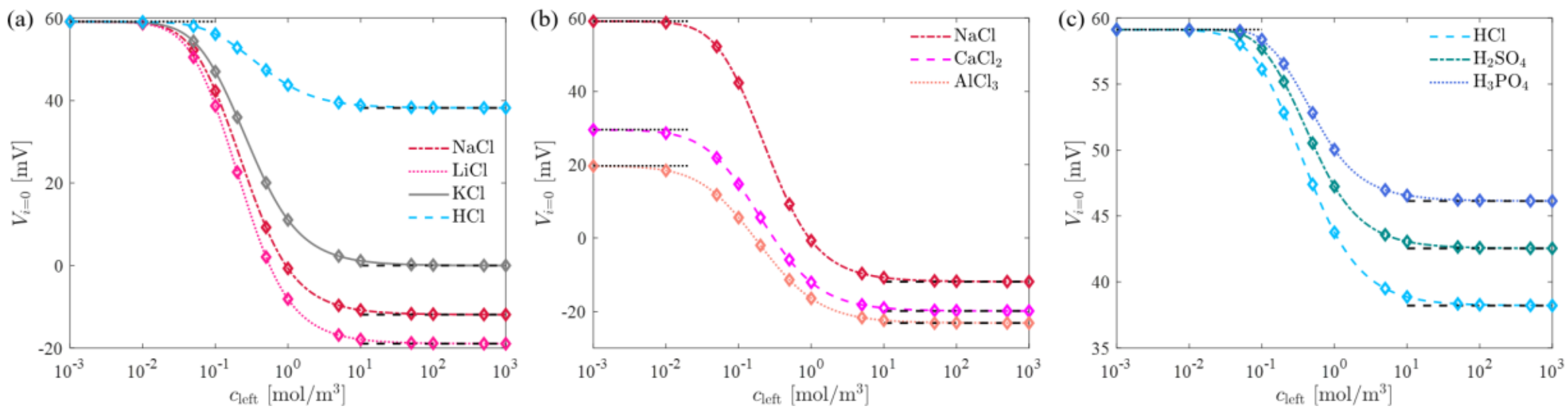

**Figure 2.** The zero-current voltage ($V_{i=0}$) versus the left bulk concentration ($c_{\text{left}}$) when $c_{\text{right}} / c_{\text{left}} = 10$ for (a) symmetric electrolytes, $z_+ = -z_-$, (b) $z_- = 1$ and varying $z_+$, and (c) $z_+ = 1$ and varying $z_-$. The colored curves are given by Eq. (22). The high and low concentration [Eqs. (24) and (25), respectively], are given by dashed and dotted black lines, respectively, while markers denote Comsol simulation data points. Simulation parameters are given in Table 1 and Table 2 in Appendix A.

Figure 2(a) focuses on the particular scenario that $z_+ = -z_- = 1$ [Eq. (23)], but with varying ratios of $D_+ / D_-$. The analytical model holds for all concentrations. However, the results at low and high concentrations are immensely interesting. At low concentrations, when the system is highly selective, all solutions converge to the same value (dotted black lines), given by Eq. (25), which is independent of the diffusion coefficients. At high concentrations, the solutions converge to Henderson's equation [Eq. (24)] denoted by the dashed black lines, where the results strongly depend on the ratio $D_+ / D_-$ (or the difference $D_+ - D_-$). For KCl ($D_+ = D_-$), we find the expected result that $V_{i=0} = 0$. For HCl ($D_+ > D_-$), $V_{i=0} > 0$ and does not change sign relative to Eq. (25), whereas for NaCl and LiCl ($D_+ < D_-$), $V_{i=0} < 0$ and has changed signs relative to Eq. (25). From this figure, one can see that there is a critical concentration wherein $V_{i=0} = 0$ (see Refs. [43] for more details on how this affects the harvestable energy).

Figure 2(b) shows the behavior of $V_{i=0}$, given by Eq. (22), for set $z_- = -1$ and $D_-$, but with varying values for $z_+$ and different values of $D_+$ (that is always smaller than $D_-$). Here, too, the theoretical model holds for all concentrations. However, as previously mentioned, the low concentration limit [Eq. (25)] depends on $z_+$ such that the salts with higher $z_+$ have a lower $V_{i=0}$. With the sole exception that Eq. (25) now depends on $z_+$, all the results of the previous analysis hold. In particular, the high concentration results converge to Henderson's equations [Eq. (24)].

Figure 2(c) is the complementary figure to Figure 2(b), where $z_+ = 1$ and $D_+$ are fixed, while several values for $z_-$ and $D_-$ are considered (with $D_+ > D_-$). Since $z_+ = 1$ remains unchanged, the low concentration limit does not vary. Here, too, the high concentration value converges to Henderson's equation [Eq. (24)]. However, since $D_+ > D_-$, as expected, $V_{i=0}$ does not exhibit a sign change.

## IV. MULTISPECIES SOLUTION

Having derived the two-species solution and having established the methodology needed for the derivation, we can start the derivation process for the multispecies solution. Similar to the two-species solution, here too we require electroneutrality. However, we also need to make two modifications to the general approach. First, we analytically derive a set of equations for all the interfacial concentrations. However, these equations must be numerically evaluated. Second, we utilize an ad-hoc assumption that each of the concentrations has a linear profile. These two modifications allow us to numerically evaluate the analytically derived expressions for $\Delta\phi_{\text{nano}}$ and the Donnan potentials.

In this section, we return to the general derivation presented in Sec. II.B, wherein all the ionic species are denoted with a '*k*' subscript (and we no longer consider '+' or '–').

Before proceeding with the derivation, it is worthwhile to discuss the issue of electroneutrality in the reservoirs. One cannot set the concentration boundary conditions to be arbitrary. One can choose any value for $c_{k,\text{left}}$ and $c_{k,\text{right}}$ so long as one requires that $\sum_{k=1}^{K}(z_k c_k)=0$ in each of the two (left/right) reservoirs. Thus, each species can have a different ratio of $c_{k,\text{right}}/c_{k,\text{left}}$. This provides additional degrees of freedom to explore that were not possible to explore in a two-species solution.

## A. Derivation

### 1. *Donnan potentials*

As previously demonstrated for two species, we can now calculate the interfacial concentrations, which lead to the Donnan potentials and more. To this end, one must select any of the *K* species to be the species that will be used to define electroneutrality and more. We term the selected species with an $\ell$ and the remaining *K*-1 species are denoted with $\hat{k}$.

Electroneutrality is still defined by Eq. (10). However, if we isolate the $\ell^{\text{th}}$ species, we find that

$$\overline{c}_\ell=\frac{\Sigma_s}{z_\ell}-\frac{1}{z_\ell}\sum_{\hat{k}=1}^{K-1}z_k\overline{c}_k\ . \tag{26}$$

To find the interfacial concentrations, one must consider the following expressions of the EC potentials, which consider the $\ell$ species and the remaining *K*-1 species

$$z_\ell\mu_{\hat{k}}-z_{\hat{k}}\mu_\ell=RT\ln(c_{\hat{k}}^{z_\ell}/c_\ell^{z_{\hat{k}}})\ . \tag{27}$$

Note that this equation is the generalization of Eq. (14) shown in the two-species problem. The requirement that the potentials are continuous at the interfaces can be rewritten as

$$c_{\hat{k}}^{z_\ell}/c_\ell^{z_{\hat{k}}}=\overline{c}_{\hat{k}}^{z_\ell}/\overline{c}_\ell^{z_{\hat{k}}}\Rightarrow\overline{c}_{\hat{k}}=c_{\hat{k}}(\overline{c}_\ell/c_\ell)^{z_{\hat{k}}/z_\ell}\ . \tag{28}$$

We can now observe that at each interface ($x=0,L$), we have *K* equations [Eqs. (26) and (28) ] that determine the *K* interfacial concentrations. This set of equations can be solved using a simple Newton-Raphson solver (which we provide in the SM [50]).

Once these equations have been evaluated numerically, one can calculate the Donnan potential by considering the $\ell^{\text{th}}$ species

$$\Delta\varphi_{\text{Donnan,left}}+\Delta\varphi_{\text{Donnan,right}}=\frac{V_{\text{th}}}{z_\ell}\ln\left(\frac{c_{\ell,\text{left}}}{c_{\ell,\text{right}}}\frac{\overline{c}_{\ell,\text{right}}}{\overline{c}_{\ell,\text{left}}}\right). \tag{29}$$

Note that this equation has the same form as the sum of Eqs. (12) and (13) shown in the two-species solution above.

### 2. *The nanochannel potential drop,* $\Delta\phi_{\text{nano}}$

In this case, the electrical current, which is determined by all the ionic species, is given by

$$\frac{i}{F}=\sum_{k=1}^{K}z_k j_k\ . \tag{30}$$

Similar to Eq. (18) from the two-species problem, upon inserting (1) into (30) and requiring that $i = 0$ yields an expression for $\partial_x \phi$

$$\partial_x \phi = -V_{\text{th}} \frac{\sum_{k=1}^{K} z_k D_k \partial_x c_k}{\sum_{k=1}^{K} D_k z_k^2 c_k} . \tag{31}$$

In principle, this equation is not integrable, not even under the assumption of local electroneutrality. However, this can be bypassed using an alternative approach that utilizes an ad hoc assumption regarding the concentration profiles inside the channel/membrane. Refs. [38–40,54,55] suggested that the concentration profiles within the channel are linear

$$\overline{c}_k(x) = \frac{\Delta \overline{c}_k}{L} x + \overline{c}_{k,\text{left}} , \tag{32}$$

where

$$\Delta \overline{c}_k = \overline{c}_{k,\text{right}} - \overline{c}_{k,\text{left}} . \tag{33}$$

Since we have already evaluated the interfacial concentrations at the two ends of the channel, all these terms are known (numerically), and integration of Eq. (31) inside the channel is now possible. Inserting Eq. (32) into Eq. (31) and integrating between the two inner ends of the channel yields

$$\Delta \phi_{\text{nano}} = -V_{\text{th}} \frac{\sum_{k=1}^{K} z_k D_k \Delta \overline{c}_k}{\sum_{k=1}^{K} z_k^2 D_k \Delta \overline{c}_k} \ln \left[ \frac{\sum_{k=1}^{K} z_k^2 D_k \overline{c}_{k,\text{right}}}{\sum_{k=1}^{K} z_k^2 D_k \overline{c}_{k,\text{left}}} \right] . \tag{34}$$

This is almost the same expression derived in Refs. [38–40], albeit with some differences. In Ref. [39], the author assumed that $\Sigma_s = 0$. Under such an assumption, the Donnan potential drops are identically zero, which means that the interfacial concentrations are always equal to their bulk values, which yields incorrect results for the interfacial concentration for any finite value of $\Sigma_s$. This is similar to Hickman's result [38]. However, unlike Morf's derivation [39], which is similar to the one presented in this work, Hickman's derivation process differs drastically from the one presented in this work. For the sake of completion, a thorough discussion on Hickman's work [38] and Morf's work [39] is provided in the SM [50]. In contrast, Lanteri et al. [40], whose derivation is similar to this work, no longer assumed $\Sigma_s = 0$ . Instead, they assume $c_{k,\text{right}} / c_{k,\text{left}} = c_{\text{right}} / c_{\text{left}} = 0.5$, such that all species are driven in the same direction at the same ratio. It should be noted that not only is their final expression limited to a very particular concentration gradient, but also the equations they provide for the interfacial concentrations [our Eqs. (26) and (28)], were limited to KCl/$CaCl_2$-like mixture. Our work differs from all three works [38–40], in that, throughout the derivation process, we have removed the unneeded constraints of $\Sigma_s = 0$, $c_{k,\text{right}} / c_{k,\text{left}} = 0.5$, and generalized the conditions needed to calculate the internal interfacial concentrations.

### 3. *$V_{i=0}$ for multispecies*

Finally, upon inserting Eqs. (29) and (34) into Eq. (5), we find that the transmembrane potential is given by

$$V_{i=0}=\frac{V_{\text{th}}}{z_\ell}\ln\left(\frac{c_{\ell,\text{right}}}{c_{\ell,\text{left}}}\frac{\bar{c}_{\ell,\text{left}}}{\bar{c}_{\ell,\text{right}}}\right)+V_{\text{th}}\frac{\sum_{k=1}^{K}z_kD_k\Delta\bar{c}_k}{\sum_{k=1}^{K}z_k^2D_k\Delta\bar{c}_k}\ln\left[\frac{\sum_{k=1}^{K}z_k^2D_k\bar{c}_{k,\text{right}}}{\sum_{k=1}^{K}z_k^2D_k\bar{c}_{k,\text{left}}}\right], \tag{35}$$

where the inner concentrations have been calculated via Eqs. (26) and (28).

## B. Reduction to previous models

Equation (35) for multispecies electrolytes is the second key finding of this work. Similar to the two-species solution analysis conducted in Sec. III.B, before proceeding with the analysis of the behavior of Eq. (35), it is beneficial to note that this multispecies model, in appropriate limits, can be reduced to several previously derived models. This implies that even though Eq. (35) utilizes the ad hoc assumption of a linear concentration profile, it is their unifying model.

*Two species*. For the case of only two species, $\ell$ and $\hat{k}$, and upon using electroneutrality, $z_\ell\bar{c}_\ell=\Sigma_s-z_{\hat{k}}\bar{c}_{\hat{k}}$, we obtain that $c_{\ell,\text{left/right}}=c_{\text{left/right}}/|z_\ell|$ and $z_\ell\Delta\bar{c}_\ell=-z_{\hat{k}}\Delta\bar{c}_{\hat{k}}$, such that Eq. (35) reduces to

$$V_{i=0}=\frac{V_{\text{th}}}{z_\ell}\ln\left(\frac{c_{\text{right}}}{c_{\text{left}}}\frac{\bar{c}_{\ell,\text{left}}}{\bar{c}_{\ell,\text{right}}}\right)+V_{\text{th}}\frac{D_\ell-D_{\hat{k}}}{z_\ell D_\ell-z_{\hat{k}}D_{\hat{k}}}\ln\left[\frac{z_{\hat{k}}(z_\ell D_\ell-z_{\hat{k}}D_{\hat{k}})\bar{c}_{\hat{k},\text{right}}-z_\ell D_\ell\Sigma_s}{z_{\hat{k}}(z_\ell D_\ell-z_{\hat{k}}D_{\hat{k}})\bar{c}_{\hat{k},\text{left}}-z_\ell D_\ell\Sigma_s}\right]. \tag{36}$$

Upon setting $\ell=+$ and $\hat{k}=-$, Eq. (36) reduces to Eq. (22), previously derived in Sec. III. This reduction implies that the key behaviors previously described for two species will also hold for multiple species. This will soon be demonstrated.

*High concentrations*. Another important limit is when the bulk concentrations are much higher than $\Sigma_s$ such that $\Sigma_s\ll c_{k,\text{left/right}}$ for all *k*. In the high concentration limit, one may set $\Sigma_s=0$, such that the channel is effectively nonselective and the inner interface concentrations converge to the bulk values ($\bar{c}_{k,\text{left/right}}=c_{k,\text{left/right}}$), implying that the Donnan potential drops vanish. Thus, Eq. (35) reduces to Henderson's equation for multispecies

$$V_{i=0}=V_{\text{th}}\frac{\sum_{k=1}^{K}z_kD_k\Delta c_k}{\sum_{k=1}^{K}z_k^2D_k\Delta c_k}\ln\left[\frac{\sum_{k=1}^{K}z_k^2D_kc_{k,\text{right}}}{\sum_{k=1}^{K}z_k^2D_kc_{k,\text{left}}}\right], \tag{37}$$

where $\Delta c_k=c_{k,\text{right}}-c_{k,\text{left}}$. It is essential to note that the sign of $V_{i=0}$ is now determined by two contributions. First, the expression $\sum_{k=1}^{K}(z_kD_k\Delta c_k)$ no longer depends just on the diffusion coefficients and valences, but also on the concentration differences. Second, the logarithmic term is no longer a simple expression that reduces to the ratio of the bulk concentrations, rather it is the ratio related to the conductivity in the two reservoirs. If a $F^2/RT$ appeared in both the numerator and denominator, these would be precisely the local conductivities in the two reservoirs.

*Low concentrations*. Derivation of the low concentration limit is trickier. Here, when the bulk concentrations are significantly lower than $\Sigma_s$ (such that $\Sigma_s\gg c_{k,\text{left/right}}$), the channel is highly selective. In this regime, co-ions (i.e., negative ions) are almost completely excluded from the channel, such that for all the <u>negative</u> ions, $\bar{c}_{k,\text{right}}=\bar{c}_{k,\text{left}}\approx 0$. Then, the concentrations of the <u>positive</u> ions depend strongly on the concentration ratio $c_{k,\text{right}}/c_{k,\text{left}}$. Namely, if all remaining positive species have the same ratio (such that $c_{k,\text{right}} / c_{k,\text{left}} = const$ for all $k$) or the ratio is different for every species (such that $c_{k,\text{right}} / c_{k,\text{left}} = const_k$ for all $k$). For the latter case, we were unable to find a reduced analytical solution. Here, one needs to use Eq. (35).

If $c_{k,\text{right}} / c_{k,\text{left}} = const$ for all species, then $\Delta\phi_{\text{nano}}$ [the second term in Eq. (35)] converges to zero, such that Eq. (35) reduces to what can be considered to be a multispecies generalization for the two-species Nernst potential [Eq. (25)]

$$V_{i=0} = \frac{V_{\text{th}}}{z_\ell} \ln\left( \frac{c_{\ell,\text{right}}}{c_{\ell,\text{left}}} \frac{\overline{c}_{\ell,\text{left}}}{\overline{c}_{\ell,\text{right}}} \right). \tag{38}$$

Several comments regarding Eq. (38) are warranted. Note that this expression is explicitly and implicitly independent of the diffusion coefficients. This result might initially be a bit surprising. However, if one bears in mind that the two-concentration limit [Eq. (25)], where $c_{k,\text{right}} / c_{k,\text{left}} = const$, is also independent of diffusion, and that the more general case [Eq. (35) ] is not diffusion independent, the result is no longer surprising. Also, note that for two species, the low concentration limit results in $\overline{c}_{\ell,\text{left}} / \overline{c}_{\ell,\text{right}} \approx O(1)$ such that $V_{i=0}$ depends on the concentration ratio of the two reservoirs ($c_{\ell,\text{left}} / c_{\ell,\text{right}}$). For multiple species, $V_{i=0}$ still depends on $\overline{c}_{\ell,\text{left}} / \overline{c}_{\ell,\text{right}}$ which accounts for all the remaining interfacial concentrations (that do not exist for two species). Thus, in contrast to the two-species solution, one cannot neglect the second ratio by assuming that the potential difference depends only on bulk properties. Rather, it is essential to calculate all the interfacial concentrations beforehand [and then use Eq. (35)].

### C. Results for 3 and 4 species

For simplicity, in most of the cases considered below, we will assume that $c_{k,\text{right}} / c_{k,\text{left}} = const$ is the same for all species. However, we will also consider the case when this does not hold.

#### *1. Concentration profiles*

Before we proceed with analyzing Eq. (35), it is worthwhile to demonstrate that numerical simulations entirely agree with the ad-hoc assumption of a linear concentration profile. Figure 3 and Figure 4 show the linear concentration profiles for a three- and four-species electrolyte, respectively, for $c_{k,\text{right}} / c_{k,\text{left}} = const$ for all species. We show that when $i = 0$, the concentrations are always linear, regardless of whether we are at low concentrations [Figure 3(a) and Figure 4(a)], intermediate concentrations [Figure 3(b) and Figure 4(b)], and high concentrations [Figure 3(c) and Figure 4(c)].

At low concentrations [Figure 3(a) and Figure 4(a)], it can be observed that the concentration of the negatively charged species is remarkably small (relative to the positive species. Moreover, it can be observed that the sum of the positive species balances out the excess counterion concentration $\Sigma_s$ such that electroneutrality is conserved.

At very high concentrations, [Figure 3(c) and Figure 4(c)], when $\Sigma_s / c_{\text{left,right}} \approx 0$, electroneutrality ensures that the sum of the positive and negative ions is the same. At intermediate concentrations [Figure 3(b) and Figure 4(b)], when the effects of selectivity are diminished but have not completely vanished, we see that to satisfy electroneutrality, there is an offset of the two positive species relative to the two negative species.

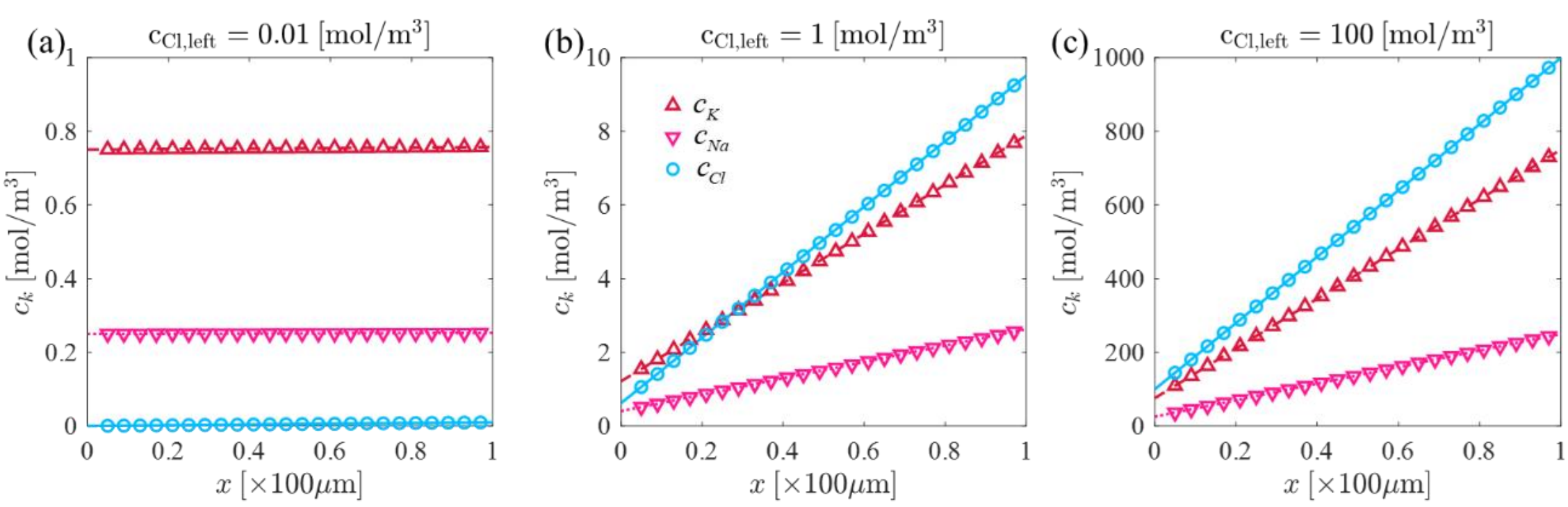

**Figure 3**. Concentration profiles for a three-species KCl/NaCl mixture for (a) low bulk concentrations, $c_{\mathrm{Cl,left}} = 0.01[\mathrm{mol/m^3}]$, (b) intermediate bulk concentrations, $c_{\mathrm{Cl,left}} = 1[\mathrm{mol/m^3}]$, and (c) high bulk concentrations, $c_{\mathrm{Cl,left}} = 100[\mathrm{mol/m^3}]$. To satisfy electroneutrality in the left reservoir, we set $c_{\mathrm{K,left}} = \frac{3}{4} c_{\mathrm{Cl,left}}$ and $c_{\mathrm{Na,left}} = \frac{1}{4} c_{\mathrm{Cl,left}}$ while in the right reservoir, we use the ratio $c_{k,\mathrm{right}} / c_{k,\mathrm{left}} = 10$ for all species. Markers denote Comsol simulations, while the solid lines are the linear profile of Eq. (32). Simulation parameters are given in Table 1 and Table 2.

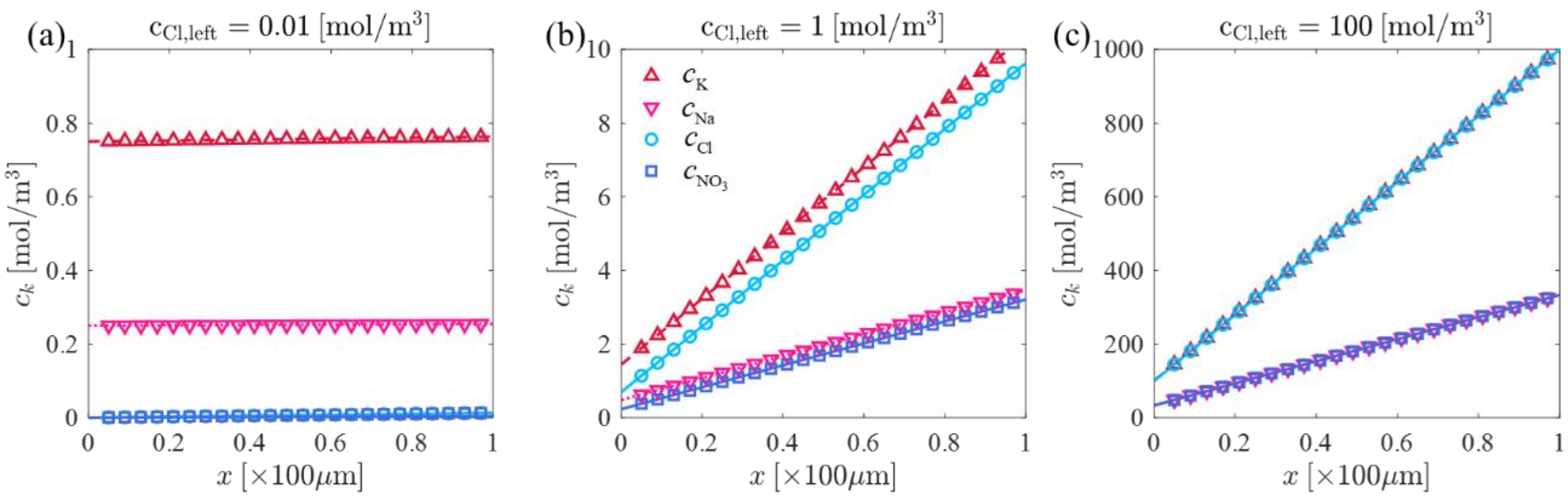

**Figure 4**. Concentration profiles for a four-species KCl/NaNO3 mixture for (a) low bulk concentrations, $c_{\mathrm{Cl,left}} = 0.01[\mathrm{mol/m^3}]$, (b) intermediate bulk concentrations, $c_{\mathrm{Cl,left}} = 1[\mathrm{mol/m^3}]$, and (c) high bulk concentrations, $c_{\mathrm{Cl,left}} = 100[\mathrm{mol/m^3}]$ . To satisfy electroneutrality in the left reservoir, we set $c_{\mathrm{K,left}} = c_{\mathrm{Cl,left}}$, $c_{\mathrm{Na,left}} = \frac{1}{3} c_{\mathrm{Cl,left}}$, and $c_{\mathrm{NO,left}} = \frac{1}{3} c_{\mathrm{Cl,left}}$ while in the right reservoir, we use the ratio $c_{k,\mathrm{right}} / c_{k,\mathrm{left}} = 10$ for all species. Markers denote Comsol simulations, while the solid lines are the linear profile of Eq. (32). Simulation parameters are given in Table 1 and Table 2.

Figure 5 shows the concentration profiles for the cases that $c_{k,\mathrm{right}} / c_{k,\mathrm{left}}$ is not the same for the three species of (K, Na, Cl) and (H, Na, Cl). In principle, all the previously described behaviors (selectivity and electroneutrality) in Figure 3 and Figure 4 are repeated here as well, such that we do not repeat the analysis. However, there is one difference worth describing. In high and intermediate concentrations [Figure 5(b),(c),(e), (f)], the profiles appear to be linear and captured well by Eq. (32). In contrast, for low concentrations [Figure 5(a),(d)], it appears that the profiles are very close to being linear. However, they also exhibit some curvature, becoming more apparent as the difference between the diffusion coefficients of the positive species increases (H vs. K).

The origin of the curvature is not yet fully understood, and it is essential to remember that the assumption of a linear profile [Eq. (32)] is still an ansatz that lacks, at this point, a rigorous mathematical proof and justification. Our exploration of a large number of electrolytes (different valences and diffusion coefficients) and a wide range of concentrations and concentration ratios did not yield an example where the linear profile failed. Whether we were "fortunate" or whether the assumption is, in fact, universal, remains to be determined and should be the focus of future works. However, in all the scenarios considered in this work, numerical simulations of $V_{i=0}$ did not deviate from Eq. (35). As such, under the assumption of a linear profile, one should treat Eq. (35) as a "convenient expression" that should always be questioned.

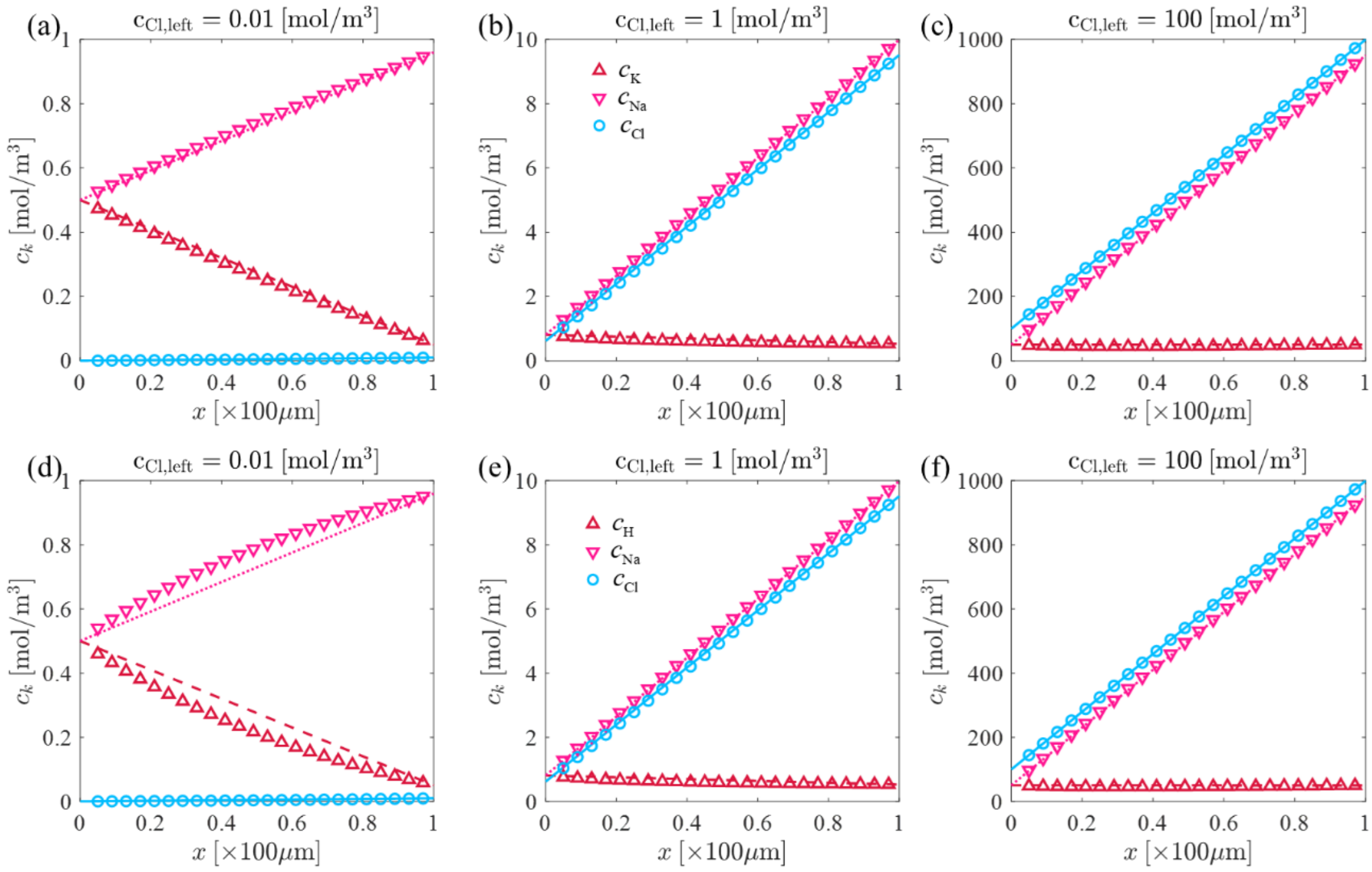

**Figure 5**. Concentration profiles for a three-species mixture, KCl/NaCl (top row) and HCl/NaCl (bottom row), for (a, d) low bulk concentrations, $c_{\text{Cl,left}} = 0.01[\text{mol/m}^3]$, (b, e) intermediate bulk concentrations, $c_{\text{Cl,left}} = 1[\text{mol/m}^3]$, and (c, f) high bulk concentrations, $c_{\text{Cl,left}} = 100[\text{mol/m}^3]$. To satisfy electroneutrality in the left reservoir, we set $c_{\text{K/H,left}} = c_{\text{Na,left}} = \frac{1}{2} c_{\text{Cl,left}}$ while in the right reservoir, we use the ratio $c_{\text{K/H,right}} / c_{\text{K/H,left}} = 1$, $c_{\text{Na,right}} / c_{\text{Na,left}} = 19$, and $c_{\text{Cl,right}} / c_{\text{Cl,left}} = 10$. Markers denote Comsol simulations, while the solid lines are the linear profile of Eq. (32). Simulation parameters are given in Table 1 and Table 2.

### 2. $V_{i=0}$

Figure 6 and Figure 7 present $V_{i=0}$ for three and four-species electrolyte solutions, respectively, when $c_{k,\text{right}} / c_{k,\text{left}} = const$ for all species for a wide range of electrolytes. Figure 8 presents $V_{i=0}$ for a particular three and four-species electrolyte solutions but varying ratios of $c_{k,\text{right}} / c_{k,\text{left}}$. In all the presented plots, we compare numerical simulations to our theoretical model given by Eq. (35) where we can see remarkable correspondence for all the considered electrolytes, concentrations, and concentration ratios.

In particular, we see that at high concentrations, the solution always converges to the "generalized" Henderson equation given by Eq. (37), which depends on the diffusion coefficients and valences, and the two local conductivities. Consider Figure 6(a), which has KCl and another species. For KCl, we saw that $V_{i=0}=0$ [Figure 2(a)]. Now, we see that $V_{i=0}\neq 0$ where the value (and its sign) now depends on the third salt. For H, $V_{i=0}>0$, while for Na and Li $V_{i=0}<0$. In all three considered cases, the signs are the same, but the absolute values are smaller where the "weighted contribution" of each salt is determined by the terms in Eq. (37). This behavior is repeated for the salts considered in Figure 6(b)-(c). The convergence to the "generalized" Henderson equation given by Eqs. (37) can also be observed in Figure 7 and Figure 8, however, an intuitive comparison to two species is no longer possible.

At low concentrations, when $c_{k,\text{right}}/c_{k,\text{left}}=const$, we see that the values converge to the generalized "Nernst" potential given by Eq. (38). Our results show that the low concentration value is primarily determined by $c_{k,\text{right}}/c_{k,\text{left}}=const$ where the numerical values converge to those given in Figure 2 [and/or Eq. (25)]. However, in contrast to the two-species case where $V_{i=0}$ varies from the Nernst potential to the Henderson potential monotonically, in three and four species, the change is no longer monotonic [Figure 6(b)]. This, of course, is related to the much more complicated expressions embedded into Eq. (35) that now depend on multiple terms and ratios that depend on multiple valences and diffusion coefficients.

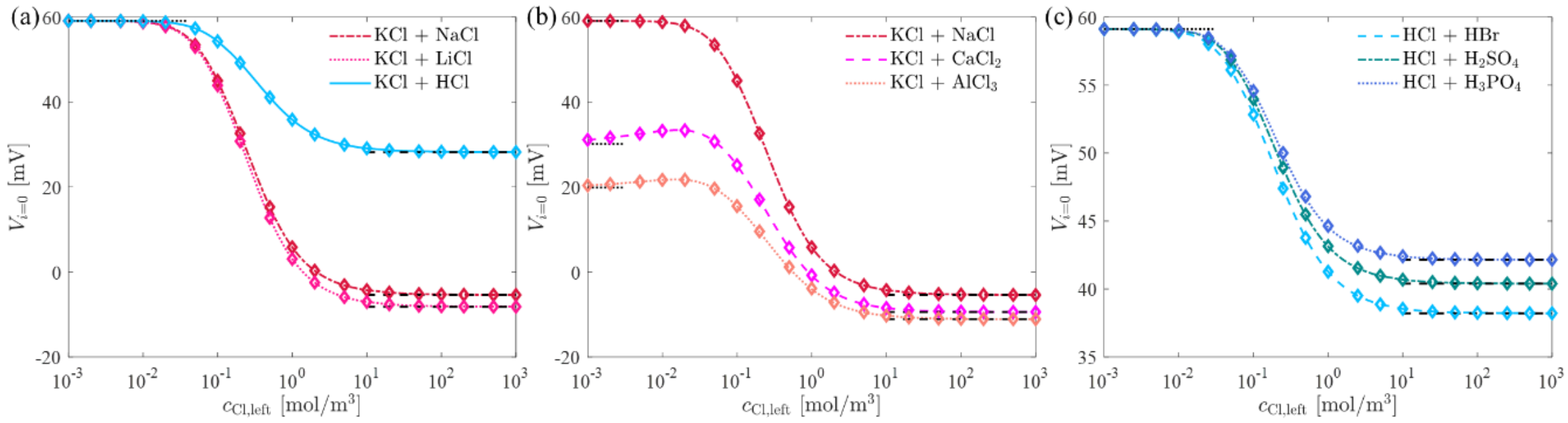

**Figure 6.** The zero-current voltage ($V_{i=0}$) versus the left bulk concentration of Cl (i.e., $c_{\text{Cl,left}}$) for various three-species electrolyte mixtures when $c_{k,\text{right}}/c_{k,\text{left}}=10$ for all ($K$-)species. In (a), a Cl-based salt with a valency of 1 is added to KCl where $c_{\text{K,left}}=\frac{1}{2}c_{\text{Cl,left}}$ and $c_{\text{S,left}}=\frac{1}{2}c_{\text{Cl,left}}$ for S=Na, Li, H. In (b), a Cl-based salt with varying cation valences (1,2, and 3) is added to KCl where $c_{\text{K,left}}=\frac{1}{2}c_{\text{Cl,left}}$ and $c_{\text{S,left}}=\frac{1}{2}(c_{\text{Cl,left}}/z_S)$ for S=Na, Ca, Al. In (c), there is only one positive ion (H), while the negative ions are Cl and another ion with varying valency, where $c_{\text{H,left}}=2c_{\text{Cl,left}}$ and $c_{\text{S,left}}=-c_{\text{Cl,left}}/z_S$ for S=Br, SO4, PO4. The colored curves are given by Eq. (35). The high and low concentration limits [Eqs. (37) and (38), respectively], are given by dashed and dotted black lines, while markers denote Comsol simulation data points. Simulation parameters are given in Table 1 and Table 2.

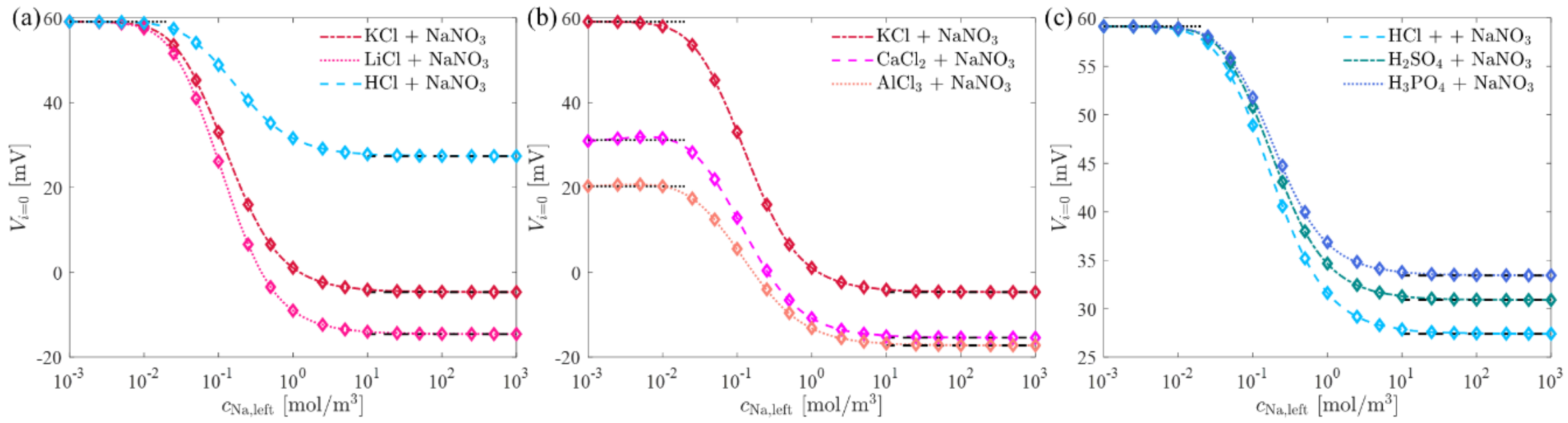

**Figure 7**. The zero-current voltage ($V_{i=0}$) versus the left bulk concentration of Na (i.e., $c_{\mathrm{Na,left}}$) for various four-species electrolyte mixtures when $c_{k,\mathrm{right}} / c_{k,\mathrm{left}} = 10$ for all ($K$-)species. In (a), a Cl-based salt with a valency of 1 is added to $NaNO_3$, where $c_{\mathrm{Cl,left}} = c_{\mathrm{NO_3,left}} = c_{\mathrm{Na,left}}$ and $c_{\mathrm{S,left}} = c_{\mathrm{Na,left}}$ for S=K, Li, H. In (b), a Cl-based salt with varying cation valences (1,2, and 3) is added to $NaNO_3$, where $c_{\mathrm{Cl,left}} = c_{\mathrm{NO_3,left}} = c_{\mathrm{Na,left}}$ and $c_{\mathrm{S,left}} = c_{\mathrm{Na,left}} / z_S$ for S=K, Ca, Al. In (c), salt with H and varying anion valences (-1,-2, and -3) is added to $NaNO_3$, where $c_{\mathrm{H,left}} = c_{\mathrm{NO_3,left}} = c_{\mathrm{Na,left}}$ and $c_{\mathrm{S,left}} = -c_{\mathrm{Na,left}} / z_S$ for S=Cl, $SO_4$, $PO_4$. The colored curves are given by Eq. (35). The high and low concentration [Eqs. (37) and (38), respectively], are given by dashed and dotted black lines, while markers denote Comsol simulation data points. Simulation parameters are given in Table 1 and Table 2.

Figure 8(a-c) present $V_{i=0}$ for KCl+NaCl, HCl+NaCl, and KCl+$NaO_3$ solutions, respectively, where we vary the ratio of $c_{k,\mathrm{right}} / c_{k,\mathrm{left}}$. Importantly, we see that even though the concentration profiles in Figure 5 were not exactly linear, the final result of $V_{i=0}$ given by Eq. (35) still shows streaking correspondence with numerical simulations for all concentrations, electrolytes, etc. At high concentrations, the results converge to the Henderson solution [Eq. (37)], while at low concentration, Eq. (38) (given by the black dashed line) fails when $c_{k,\mathrm{right}} / c_{k,\mathrm{left}}$ is not a constant. This is because, in contrast to $V_{i=0}$ presented in Figure 6 and Figure 7 (which had the value of $c_{k,\mathrm{right}} / c_{k,\mathrm{left}}$ and thus the same numerical value of $V_{i=0}$) here $V_{i=0}$ varies as $c_{k,\mathrm{right}} / c_{k,\mathrm{left}}$ is varied. This is because $V_{i=0}$ is no longer given by a simple expression such as Eq. (38), which is independent of the diffusion coefficients, and rather is given by Eq. (35), such that $V_{i=0}$ is a function of all the $c_{k,\mathrm{right}} / c_{k,\mathrm{left}}$ ratios as well as the diffusion coefficients (and valences).

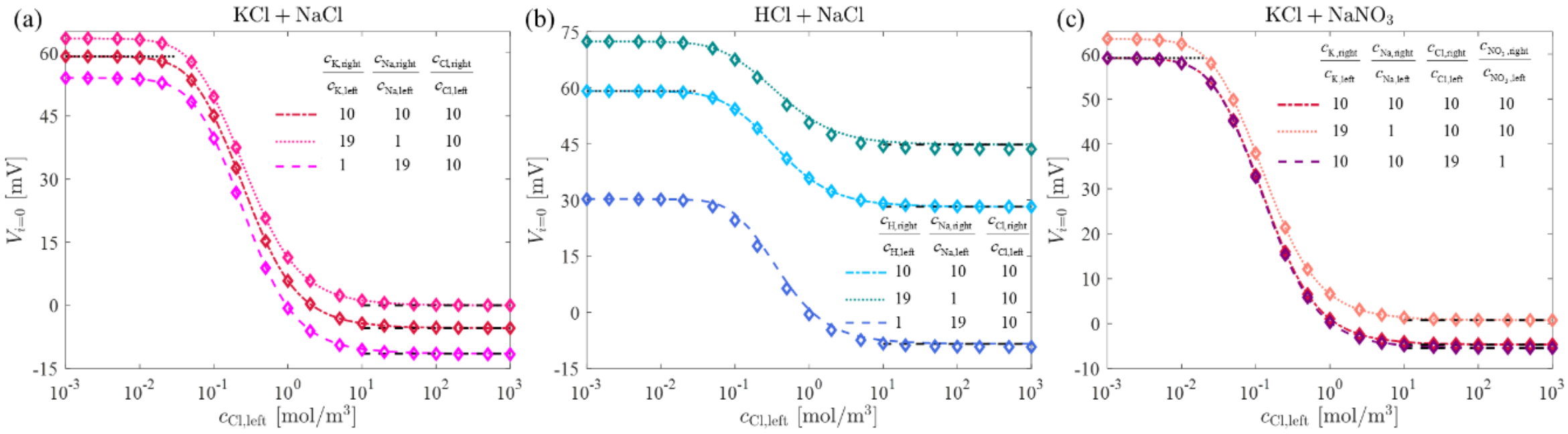

**Figure 8**. The zero-current voltage ($V_{i=0}$) versus the left bulk concentration of Cl (i.e., $c_{\mathrm{Cl,left}}$) for (a) a three-species KCl+NaCl mixture with $c_{\mathrm{K/Na,left}} = \frac{1}{2} c_{\mathrm{Cl,left}}$, $c_{\mathrm{Cl,right}} / c_{\mathrm{Cl,left}} = 10$ and varying ratios ($c_{k,\mathrm{right}} / c_{k,\mathrm{left}}$) for K and Na species, (b) a three-species HCl+NaCl mixture with $c_{\mathrm{H/Na,left}} = \frac{1}{2} c_{\mathrm{Cl,left}}$, $c_{\mathrm{Cl,right}} / c_{\mathrm{Cl,left}} = 10$ and varying ratios ($c_{k,\mathrm{right}} / c_{k,\mathrm{left}}$) for H and Na species, and (c) a four-species KCl+$NaNO_3$ mixture with $c_{\mathrm{K/Na/NO_3,left}} = c_{\mathrm{Cl,left}}$ and varying ratios ($c_{k,\mathrm{right}} / c_{k,\mathrm{left}}$). The colored curves are given by Eq. (35). The high concentration [Eqs. (37)] is given by dashed black lines, while the low concentration [Eq. (38)] is given by dotted black lines only for the case where $c_{k,\mathrm{right}} / c_{k,\mathrm{left}} = const$. Markers denote

Comsol simulation data points. Simulation parameters are given in Table 1 and Table 2.

## V. DISCUSSION, FUTURE DIRECTIONS, AND CONCLUSIONS

In this work, we have derived two analytical expressions for one of the key characteristics of ion transport systems – the transmembrane potential $V_{i=0}$ for two-species [Eq. (22)] and multispecies [Eq. (35)] electrolytes with arbitrary properties ($D_k, z_k$) subjected to an arbitrary concentration gradient ($c_{k,\text{right}} / c_{k,\text{left}}$). For two-species, the solution is exact in that an additional assumption, such as the linear concentration profile used for multispecies, is not needed.

Each of these derived expressions is then analyzed. We show that, at the appropriate limits, the two-species model [Eq. (22)] reduces to several previously derived models that are limited to symmetric $D_k$ and anti-symmetric $z_k$, and/or to the Henderson [Eq. (24)] and Nernst [Eq. (25)] potentials at high and low concentrations, respectively. A similar analysis is then provided for the multispecies solution [Eq. (35)], where we show that it can be reduced to two species, as well as to a generalized Henderson [Eq. (37)] and generalized Nernst potential [Eqs. (35) or (38)]. Our analysis demonstrates that an interplay of $D_k$ and $z_k$, as well $c_{k,\text{right}}$ and $c_{k,\text{left}}$ determine the response of $V_{i=0}$. Importantly, to ensure that our derivation process is self-consistent, including the use of the linear concentration profile assumption, we validate all theoretical results with numerical simulations. These simulations show remarkable correspondence and thus confirm our results.

From the fundamental perspective, this work serves as a starting point for several new directions. In this work, we have derived the expression for $V_{i=0}$. However, there are two additional characteristics of interest when $i = 0$. These are the electrical conductance and transport numbers. It is our hope that we will be able to utilize the method used in this work to generalize the expressions for the electrical conductance and transport numbers recently derived for two species with symmetric $D_k$ and anti-symmetric $z_k$ in Refs. [43,45].

In this work, we have ignored several additional mechanisms that are known to influence ion transport through charge-selective materials. It is our hope that these mechanisms will eventually be included into the theory for $V_{i=0}$. Of particular importance are surface charge regulation, advection, and the effects of the reservoirs. Surface charge regulation is an established mechanism [56–59] in which the adsorption of ions from the bulk, or the electrolyte, onto the surface modifies (or "regulates") the surface charge density. However, existing models for surface charge regulation generally assume that $c_{\text{left}} = c_{\text{right}}$. We hope that, eventually, the effects of surface charge regulation will also be able to be embedded into $V_{i=0}$. Similarly, to date, analytical works for $V_{i=0}$ have neglected advection. Even though the electrical current is zero, the electric field is not, and a resultant electroosmotic flow should appear. It is not clear a priorily whether the addition of advection will change the result. However, it is clear that advection should be considered. Whether theory will be able to resolve this remains to be determined. But numerical simulations will likely be able to provide valuable insights [60,61].

We also note that in this work, we have explicitly assumed that the effects of the reservoirs are negligible. However, in a submitted work [62], we derived an expression for $V_{i=0}$ for a KCl-like salt ($D_+ = D_-$ and $z_+ = -z_-$). There, we show that at very high and very low concentrations, $V_{i=0}$ reduces to the Henderson and Nernst equations [Eqs. (24) and (25), respectively], but at intermediate concentrations, the geometry of the reservoirs (whether 1D, 2D, or 3D) induces changes relative to the response that neglects the reservoirs. The merging of the results of this work and that of Ref. [62] will need to be merged for multispecies and arbitrary properties ($D_k, z_k$).

There is one last issue to discuss that lies at the interface of fundamental and application. In the electrophysiological community, the gold standard for $V_{i=0}$ is given by what is commonly known as the Goldman-Hodgkin-Katz (GHK) theory [36,37,63], given here for a two-species $z_+ = -z_-$ electrolyte {see the work of Goldman [36] for the general expression for multispecies and arbitrary properties ($D_k, z_k$)}

$$V_{\mathrm{GHK}} = V_{\mathrm{th}} \ln\left(\frac{D_+ c_{\mathrm{right}} + D_- c_{\mathrm{left}}}{D_+ c_{\mathrm{left}} + D_- c_{\mathrm{right}}}\right). \tag{39}$$

It is essential to understand that the GHK theorem is derived from the same governing equations [Eqs. (1)-(4)] used here (this was demonstrated in our previous reports [43,45,64]). The difference between GHK and the approach used in this work (whether for two species or multiple species) is in how to simplify the PNP equations.

In the GHK theorem (where the number of species is not as important as it is here), the Poisson equation [Eq. (2)] is reduced by assuming that $\partial_{xx}\phi = 0$. This leads to a linear potential or the famous uniform/constant electric field assumption. Thereafter, one can calculate the concentrations (which at high concentrations turn out to be exponential). However, the problem with this equation is that it does not satisfy local electroneutrality [Eq. (10)] or even global electroneutrality. Moreover, within GHK, one has to assume that the effects of $\Sigma_s \to 0$, such that, in all practicality, the theory is limited only to high concentrations (in fact, Goldman explicitly assumes that $\Sigma_s = 0$).

This work takes an alternative approach by reducing the Poisson equation [Eq. (2)] with electroneutrality ($\rho_e = 0$). In our two-species formulation (Sec. III), this is explicitly embedded, while in our multispecies formulation, it is implicitly embedded into the boundary conditions. Naturally, this leads to a completely different result, where it is clear that the high concentration limits are different. For two species, it is immediate to see that Eq. (39) and (24) differ and provide drastically different predictions for $V_{i=0}$ (see our recent work [43] for a comparison between these two expressions). The advantage of the approach used in this work is that, in contrast to GHK, it is not limited to high concentrations, but rather it can be extended to arbitrarily low concentrations.

One cannot end this comparison without noting an interesting mathematical peculiarity common to both GHK and this work for multiple species. In principle, the equations for the K-species are non-integrable without adding an explicit assumption. GHK assumes that the electric field is constant and then calculates the (exponential) concentration profiles. Here, we assume that the concentration profiles are linear such that the electric field is logarithmic. It is immensely interesting to see how the non-commutative assumption of linearity of either the concentration or the electric field can change the results in such a drastic manner.

An added advantage of the model presented here is that in contrast to GHK, which often translates the diffusion coefficients into permeabilities that are then taken as fitting parameters that vary across system/sample/experiments, our final result is still given in terms of the diffusion coefficients, which are not (explicitly or implicitly) fitting parameters. It is our hope that Eq. (35) [along with the boundary conditions Eqs. (26) and (28)] will find its way into the realm of electrophysiology, and it can serve as an alternative model for data interpretation.

Finally, we conclude with a few comments regarding pure applications. It is rather obvious that this new model has immediate applications for water desalination and energy harvesting, which utilize charged nanoporous membranes for filtration. However, it is our hope that this model can also serve as a first-principle model for ion-ion separation processes, such as Lithium extraction. Our new model is the first to provide a comprehensive expression accounting for many of the system parameters, and it can be used to guide and enhance the analysis and data interpretation of all these experimental systems.

*Acknowledgements***.** This work was supported by the Israel Science Foundation (Grant No. 204/25). We also acknowledge the support of the Ilse Katz Institute for Nanoscale Science & Technology and the Pearlstone Center for Aeronautical Engineering Studies. We are grateful to the anonymous referee for directing us to Refs. [38–40], which were previously unknown to us. These works prompted a substantial revision and restructuring of our manuscript, shifting its focus from a two-species framework to the considerably more challenging multispecies problem. Although this required significant additional effort, we are deeply indebted to the referee for their insightful and constructive guidance.

*Data availability***.** The data that support the findings of this article are available from the authors upon reasonable request.

## VI. APPENDIX A: NUMERICAL SIMULATIONS

We compare our theoretical results [Eq. (22) and (35)] to the numerical simulations of the 1D Poisson–Nernst–Planck equations [Eqs. (1)-(2)], subject to the boundary conditions of Eq. (4), with the simulation parameters given in Table 1 for the salts whose details are given in Table 2.

The simulations are implemented in Comsol using the Electrostatic and Transport of Diluted Species modules, as described in detail in our previous works [42,43,58,65], with the three differences. First, we no longer assume that $z_+ = -z_-$. Accordingly, we modify the boundary conditions for the bulk concentrations to account for their dependency on the valency. Second, we also consider different diffusion coefficients than those used in previous works. In Ref. [43], we considered KCl, NaCl, LiCl, and HCl. Here, we consider multi-valent salts (Table 2). Third, in previous simulations, we considered only two species. Here, we consider multiple species. To that end, we add additional conservations laws (Nernst-Planck) equations for the relevant species, and we modify the space charge density, $\rho_e$ [Eq. (2)], accordingly. The remainder of the simulation remains unchanged.

It is essential to emphasize that, in contrast to the theoretical approach that utilizes the assumption of local electroneutrality [Eq. (10)], in simulations, this is not enforced, but rather it is an outcome. In particular, local electroneutrality can be attributed to the fact that the Debye length is substantially smaller than the length of the system.

To calculate $V_{i=0}$, we employ the following strategy that is based on our past works [42,43,45]. For a given configuration, we apply a voltage difference ($V$), as defined in Eq. (4), and calculate the resultant current, $i$. We scan a wide range of voltages such that we have a detailed $i-V$ curve. We then interpolate this $i-V$ curve to find where the current is zero.

**Table 1**: Simulation parameters used for simulations. The value for $\Sigma_s$ corresponds to the approximate experimental values from Refs. [66,67].

| Parameter | Value |
|---|---|
| Excess counterion concentration, $\Sigma_s$ [mol/m$^3$] | 1 |
| Temperature, $T$ [$K$] | 298 |
| Relative permittivity, $\varepsilon_r$ | 78.4 |

**Table 2**: Aqueous electrolyte solution properties [68]

| Salt | $z_+$ | $z_-$ | $D_+[10^{-9}\ m^2/s]$ | $D_-[10^{-9}\ m^2/s]$ |
|---|---|---|---|---|
| KCl | 1 | -1 | 2.0 | 2.0 |
| NaCl | 1 | -1 | 1.33 | 2.0 |
| LiCl | 1 | -1 | 1.029 | 2.0 |
| HCl | 1 | -1 | 9.31 | 2.0 |
| HBr | 1 | -1 | 2.0 | 2.0 |
| $NaNO_3$ | 1 | -1 | 1.33 | 1.90 |
| $CaCl_2$ | 2 | -1 | 0.793 | 2.03 |
| $H_2SO_4$ | 1 | -2 | 9.31 | 1.07 |
| $AlCl_3$ | 3 | -1 | 0.559 | 2.03 |
| $H_3PO_4$ | 1 | -3 | 9.31 | 0.612 |